# Daily Affect Fluctuations in Phone Screen Content Predict Anxiety and Depressive Symptoms

Christopher A. Kelly[^,1,2], Yikun Chi[3], Nicholas Haber[4], Byron Reeves[3], Mu-Jung Cho[5], Thomas N. Robinson[3], Nilàm Ram[2,3,] & Johannes C. Eichstaedt[^,1,2,6]
[1]Institute for Human Centered AI, Stanford University
[2]Department of Psychology, Stanford University
[3]Department of Communication, Stanford University
[4]Stanford Graduate School of Education
[5]Departments of Pediatrics and Medicine, Stanford University
[6]Decision Sciences, INSEAD, France

[^]Corresponding authors: chris.kelly.stanford@gmail.com; johannes.stanford@gmail.com

**Abstract**

The relationship between digital media use and mental health remains poorly understood, in part because real-world digital behavior is rarely captured at scale. This intensive longitudinal study tracked participants' complete natural smartphone interactions over one year. We collected screenshots every 5 seconds from 145 adults (yielding 111 million screenshots), alongside biweekly assessments of anxiety and depression (mean = 24 surveys). The valence and arousal of each screenshot were assessed using a deep learning affect model. Individuals showed highly idiosyncratic media patterns, with substantially more variance in anxiety and depression accounted for within-person than between-person. Day-to-day fluctuations in the valence and arousal of a person's screen content predicted subsequent changes in depression and anxiety, whereas between-person differences did not. Specifically, greater exposure to low-arousal negative content was associated with higher depression and anxiety. These findings underscore the dynamic, idiosyncratic nature of digital consumption and the need for targeted measurement and intervention.

# Introduction

People are spending increasing amounts of time online and on smartphones (Kemp, 2023). There has been substantial research into the link between online behavior and mental health. To date, studies exploring the relationship between digital media activity and mental health have predominantly focused on the assessment of total screentime and total time spent using social media (Babic et al., 2017; Page et al., 2010; Granic et al., 2014; Odgers et al., 2018; Orben & Przybylski, 2019; Brusilovskiy et al., 2016). The findings on the impact of screentime on mental health are mixed, with some studies indicating negative associations while others report minimal, no, or even positive associations (Sanders et al., 2024). The inconsistency of findings may, in part, be due to researchers' reliance on self-reported measures of screentime and problematic social media use, which are only moderately or minimally correlated with actual screentime and use patterns (Sanders et al., 2024).

Resolution of the inconsistent findings regarding screentime and mental health likely requires both more accurate and nuanced measurement and analysis of how particular aspects of digital activity, including type of online behaviors and content consumed, may shape mental health outcomes (Kelly & Sharot, 2025; Kelly et al., 2024; Sharot & Sunstein, 2020; Orben et al., 2020). Consistent with this view, recent meta-analyses conclude that associations between social media use and mental health are typically small and heterogeneous, with more reliable links emerging for problematic use and specific patterns of engagement (e.g., active vs. passive use) than for time spent alone (Shannon et al., 2022; Hancock et al., 2022; Godard & Holtzman, 2024).

## **Affect and Mental Health**
One potentially critical feature of the content individuals consume on their screens, and one that might be related to mental health, is its affective nature. Experience of negative affect is a core part of many mental health disorders, including depression and anxiety (Watson et al., 1988; Clark & Watson, 1991). Common symptoms of many conditions include persistent negative mood states, such as sadness and irritability, and negative thoughts, including worry and rumination (Moberly & Watkins, 2008).
Given that digital activities may both shape and reflect individuals' emotional and cognitive experiences, we hypothesize that the affective content of individuals' digital environment – and centrally, what appears on their smartphone screen – intersects with their mental health.

## **Affective Content and Mental Health**
Previous research has shown that the affective content people seek online varies over time and across individuals (Kelly & Sharot, 2021; Kobayashi et al., 2019; Sunstein, 2019). These differences and changes in content may be linked to an individual's cognitive and affective states (Kelly & Sharot, 2021; Kelly et al., 2024; Kelly & Sharot, 2025; Wilding et al., 2022; Charpentier et al., 2022). For example, experiencing a negative mood may lead an individual to seek information with a similarly negative sentiment, which, in turn, may reinforce or worsen their affective state. Indeed, Kelly and Sharot (2025) examined this relationship, showing bidirectional links between the valence of online content consumption and mental health: consuming more negative content negatively impacts mental health, which then drives further engagement with negative content.

This cycle reflects the process of rumination, a repetitive and passive focus on negative thoughts and emotions that prolongs distress and deepens negative mood (Watkins et al., 2008; Michl et al., 2013). Rumination has been linked to depression, where individuals tend to seek out and engage with stimuli that reinforce their sadness (Millgram et al., 2015). Just as persistent negative thoughts can intensify low mood, persistently seeking negative content may create a similar self-reinforcing loop. In sum, there is some evidence that an individual's affective state both shapes and is shaped by the type of digital content they consume (Kelly & Sharot, 2025; Sharot & Sunstein, 2020). Understanding these patterns is critical for assessing how digital media influences mental health.

**The Gap**
Despite evidence that affective online content consumption is linked to mental health (Kelly & Sharot, 2025), critical gaps remain in our understanding. First, previous research has not assessed affective engagement across the entire digital media landscape, across all apps, not just social media and web browsing. For example, Kelly & Sharot (2025) excluded password-protected areas of digital engagement, such as social media platforms and messaging applications, leaving unanswered the question of how affective content consumed within these prevalent digital spaces influences mental health. Second, prior work has predominantly focused on textual content, overlooking the affective implications of multimedia content, which is a substantial or dominant portion of individuals' digital interactions. Finally, existing studies monitored digital behaviors over relatively short time frames, limiting insight into the long-term dynamics and stability of affective content consumption and its mental health consequences. Finally, most studies have relied on short observation windows or between-person comparisons, limiting insight into how dynamic, within-person fluctuations in digital content relate to changes in mental health over extended periods. Addressing these gaps requires a comprehensive, longitudinal, and multimodal approach that captures detailed digital interactions across the full digital ecosystem (e.g., internet browsing, messaging, social media, video platforms, gaming, dating apps, and email) to clarify how variations in affective content consumption relate to mental health outcomes.

Work using continuous screen capture has begun to clarify why prior media–mental health findings are often small or inconsistent: relationships appear to be highly individualized and better captured with intensive longitudinal methods. For example, Cerit and colleagues (2025) report strong, person-specific associations between fragmented smartphone use metrics and mental health dimensions over 1 year in five adults. These findings suggest that digital media effects may be detectable at the within-person level but may also vary substantially across individuals. We extend this measurement-first, within-person approach by quantifying the valence and arousal of on-screen content at scale and testing how within-person affective fluctuations in daily exposure relate to fortnightly depressive symptoms and anxiety across a large sample.

**Human Screenome Project: Objective Measurement of Daily Digital Media Behavior and Longitudinal Measurement of Mental Health**
To address the limitations of previous studies, this project uses data from the Human Screenome Project (Reeves et al., 2020), which obtains objective measures of individuals' digital behavior by capturing all the content that appears on participants' smartphone screens over up to a year. Unlike prior research that relied on self-reported measures or focused narrowly on specific platforms (e.g., social media or web-browsing), the Human Screenome Project observes how individuals interact with the entire digital landscape, including email, messaging, and social media applications. Alongside the image time-series data chronicling content, participants in this study completed fortnightly self-reports on depressive symptoms (CES-D; Radloff, 1977) and anxiety (STAI-Trait; Spielberger et al., 1983). Together, the two data streams enable a detailed examination of how digital content consumption patterns relate to mental health outcomes.

### *Affective Digital Footprints*

This paper introduces the concept of *Affective Digital Footprints*, formally defined as longitudinal patterns of valence and arousal characterizing the digital media content individuals consume on their smartphones. This construct operationalizes affective consumption within a multidimensional framework of emotion, emphasizing how interactions between valence (the positivity or negativity of emotional content) and arousal (the intensity or activation level of emotional content) shape affective experiences (Russell, 1980; Feldman Barrett, 2017). Using comprehensive Human Screenome Project data (spanning 111,270,108 screenshots from 145 individuals), collected intensively and continuously over extended periods, allows us to overcome previous research limitations associated with short-term monitoring and narrow content analyses. Specifically, the multidimensional, longitudinal approach captured by *Affective Digital Footprints* enables detailed examination of both stable, trait-like preferences in content consumption and dynamic, day-to-day fluctuations. This framework thus facilitates precise analysis of how within-person changes and between-person differences in digital affective content consumption are systematically associated with mental health outcomes.

### *The Present Study*

This study is guided by two key research questions.
*RQ1:* To what extent are *Affective Digital Footprints* consistent within individuals over time and distinct between individuals?
*RQ2:* Are *Affective Digital Footprints* associated with mental health outcomes, such as depression and anxiety?

To investigate these questions, we use data from the Human Screenome Project, which employs the Screenomics protocol to capture real-world digital behavior. This dataset includes 111,270,108 screenshots (median = 751,692, min = 4,204, max = 2,471,180) collected every 5 seconds while smartphones were in use and self-reported mental health assessments completed fortnightly over a year (mean number of weeks = 24.06, min = 15 , max = 26) from 145 participants (Mean Age = 45; Female = 51.4%, White = 46.6%, Black = 17.6%, Asian = 6.7%, Hispanic = 4.3%, Native American = 2.4%, Multiracial = 22.4%). Every two weeks, participants completed self-report inventories on their experience of depressive symptoms over the previous two weeks (10-item Center for Epidemiologic Studies Depression Scale [CES-D]; Radloff, 1977) and anxiety (20-item State-Trait Anxiety Inventory [STAI]; Spielberger et al., 1983). These super-intensive longitudinal data provide a unique opportunity to examine whether and how fluctuations in individuals' *Affective Digital Footprints* correspond to fortnightly changes in depressive symptoms and anxiety.

Figure 1 summarizes the analysis of screenshot data acquisition and valence and arousal extraction pipeline. The Deep Affect model (Parry & Vuong, 2021) was trained on multiple benchmark datasets (e.g., IAPS (Lang et al., 2005), OASIS (Kurdi et al., 2017)), achieving 87% accuracy for valence and 88% for arousal when compared with human ratings, and validated for application to smartphone screenshot data (Rocklin et al., 2023).

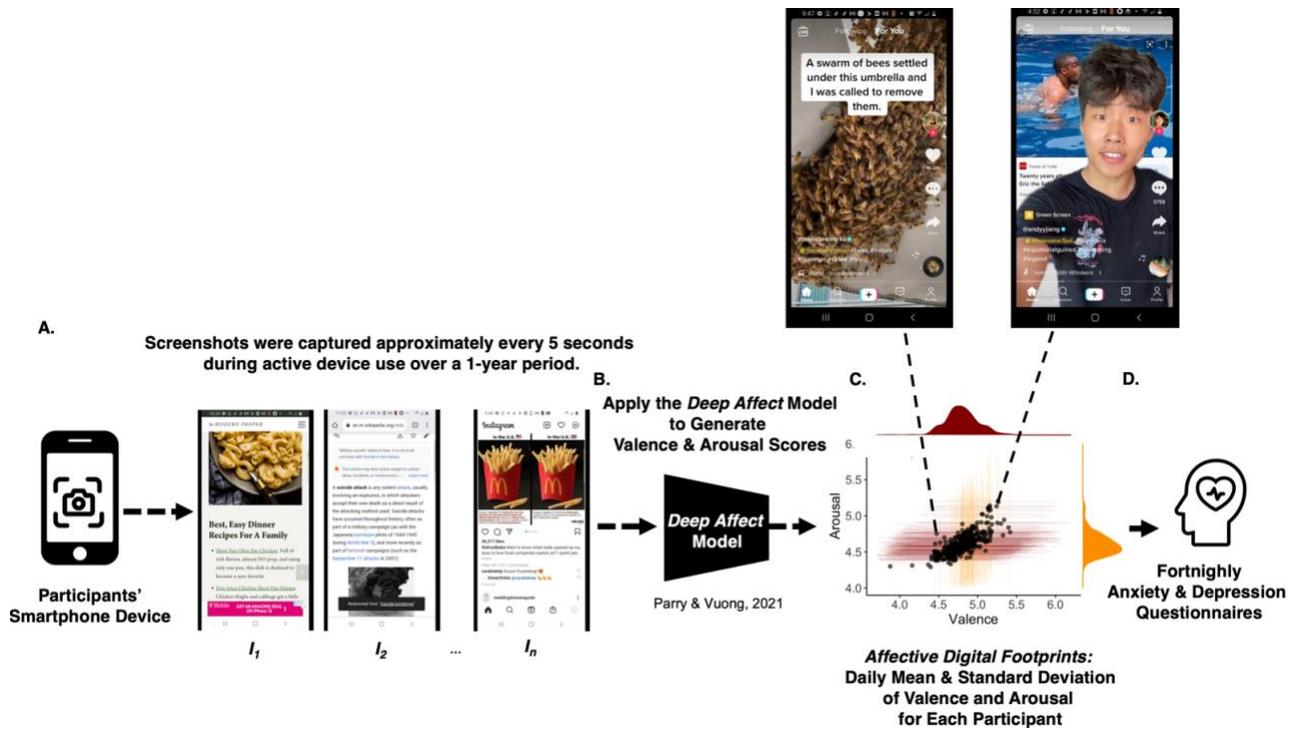

**Figure 1. Overview of the Screenomics method for capturing digital media content consumption. A.** 145 participants provided 111,270,108 screenshots approximately every five seconds during smartphone use. **B.** Each screenshot was analyzed for affect using the Deep Affect model (Parry & Vuong, 2021), which assigns continuous valence scores (1 = negative to 9 = positive) and arousal scores (1 = calm to 9 = exciting) based on the affective qualities of visual elements. **C.** A daily *Affective Digital Footprint* was then created for each participant, summarized by the daily average and standard deviation of valence and arousal scores of each day's smartphone content. Two representative screenshots are provided to exemplify the types of content contributing to the participant's daily valence and arousal scores. **D.** Daily Affective Digital Footprints are examined in relation to fortnightly self-reports of mental health (depressive symptoms and anxiety).

**Results**

***RQ1: Mapping Affective Digital Footprints: Assessing Between-Participant Differences and Within-Participant Stability.*** To calculate the affective characteristics of the content participants consume, the valence and arousal of their screenshots were calculated using the Deep Affect model (Parry & Vuong, 2021; **see Methods**). For each participant, we calculated the daily average valence and arousal scores of the screenshots, representing the affective quality of the content they consumed (**see Figure**

**1 & 2A**). Application of the Deep Affect model to smartphone screenshot data was validated using a portion of Human Screenome Project data (**see Methods**; Rocklin et al., 2023). Here, we extend the application to this study's sample.

To answer **RQ1**, we examined the stability of valence and arousal in the digital content participants consumed (i.e., their *Affective Digital Footprints*) by applying separate random intercept models for valence and arousal. These models included a random intercept for each participant to account for the hierarchical structure of the data, capturing both stable, between-participant differences in affective content consumption and within-participant variability over time.

For mean valence, the model revealed a significant fixed effect (M = 4.77, SE = 0.021, $p < 0.001$), with between-participant variance accounting for 42.6% of the total variance (ICC = 0.426, 95% CI = [0.366, 0.482]). Similarly, for mean arousal, the model showed a significant fixed effect (M = 4.58, SE = 0.013, $p < 0.001$), with between-participant variance explaining 41.0% of the total variance (ICC = 0.410, 95% CI = [0.351, 0.464]).

For the standard deviation of valence, the model revealed a significant fixed effect (M = 0.689, SE = 0.011, $p < 0.001$), with between-participant variance accounting for 48.6% of the total variance (ICC = 0.486, 95% CI = [0.426, 0.539]). Similarly, for the standard deviation of arousal, the model showed a significant fixed effect (M = 0.465, SE = 0.008, $p < 0.001$), with between-participant variance explaining 49.7% of the total variance (ICC = 0.497, 95% CI = [0.437, 0.549]).

These findings demonstrate that while individual differences in digital affective content consumption are substantial, more than half of the day-to-day variance arises from within-participant fluctuations, underscoring the dynamic nature of participants' affective digital content consumption. Overall, the ICC values suggest that *Affective Digital Footprints* capture both meaningful between-participant differences and point to the importance of within-participant variability.

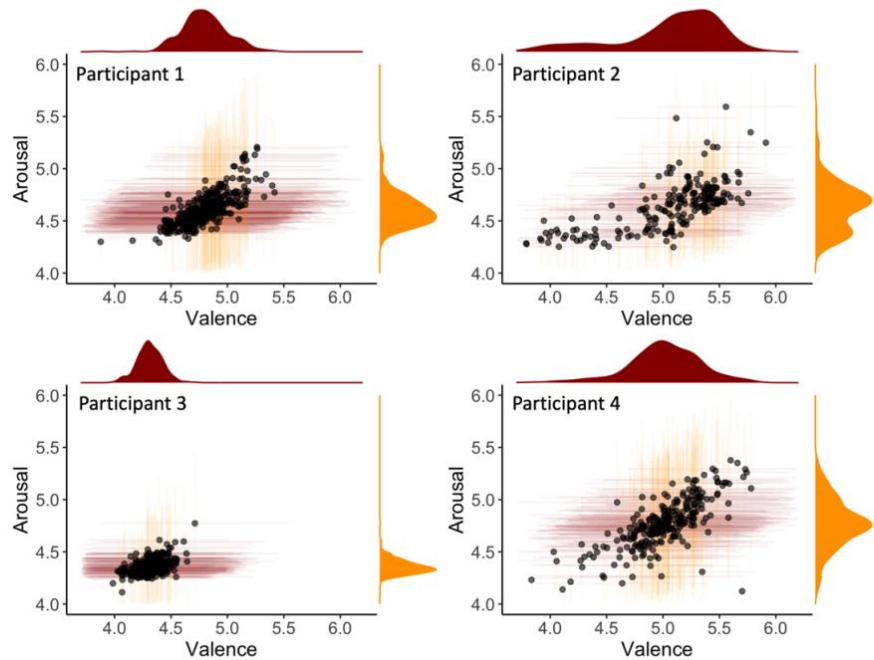

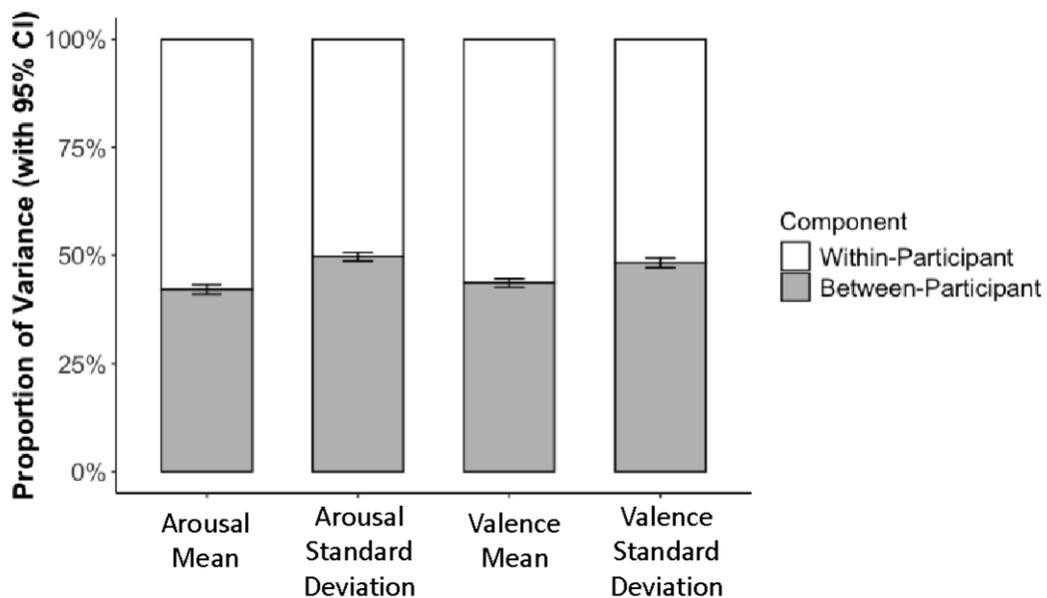

**Figure 2.** *Affective Digital Footprints* **within-person and between person variances.**
**A.** Scatter plots with marginal density distributions for valence (dark red) and arousal (orange) scores for four participants, highlighting individual variations in the affective content consumed daily. Each scatter plot represents a different participant's daily mean valence and arousal scores (shown as individual dots), with their standard deviation indicated by horizontal lines for valence (dark red) and vertical lines for arousal (orange). The density plots along each axis show the distribution of these daily mean values. The pattern and range of affective experience differs substantially across participants. **B.** Proportion of variance in daily average valence and arousal (Mean) and their variability (SD) based on an Intraclass Correlation Coefficient (ICC) analysis, distinguishing between-participant variance (grey) from within-participant variance (white). Between-participant variance represents stable differences among individuals in their content preferences (Mean) and in the variability of those preferences (SD). Conversely, within-participant variance indicates daily fluctuations in content choices and variability within individuals. Error bars indicate 95% confidence intervals for estimates. These findings suggest that individual characteristics contribute to a stable component in content preferences, though substantial day-to-day variability occurs within participants' content selection.

**RQ2: Assessing the Relationship Between *Affective Digital Footprints* and Mental Health:** Having established that participants' *Affective Digital Footprints* reflect both between-participant differences and within-participant variability, we next explored how these differences and fortnightly changes relate to mental health outcomes. We hypothesized that mental health is shaped by both stable, trait-like patterns of valence and arousal in consumed content (reflected in between-participant differences) and dynamic, state-level fluctuations in valence and arousal in consumed content (captured by within-participant changes). While it is established that affect of content consumption is related to mental health (Kelly & Sharot, 2025), the relative contributions of these between-participant and within-participant factors, particularly across the diverse digital media landscape, remain unclear. Understanding how these factors interact could provide new insights into the ways digital content consumption is associated with mental health outcomes.

The hypothesis that mental health is linked to valence and arousal in consumed content was tested using linear mixed-effects models examining the extent to which individual differences and within-person fluctuations in the four features of individuals' *Affective Digital Footprints* (i.e., valence, arousal, valence variability, and arousal variability) predicted self-reported depressive symptoms (Center for Epidemiologic Studies Depression Scale [CES-D]; Radloff, 1977) and anxiety (State-Trait Anxiety Inventory [STAI]; Spielberger et al., 1983). The models included both main effects and interactions of state (daily) and trait (overall) valence and arousal, as well as the daily variability in valence and arousal (SD), to assess how momentary fluctuations and stable patterns in affective content consumption influenced mental health outcomes. State (daily) variables were mean-centered within participants, and random intercepts and slopes were included for key state-level interactions to accommodate the nested nature of the data. Full model details are provided in the Methods section.

First, the linear mixed-effects model predicting self-reported depression (**see Table 1 for full results**) revealed a clear main effect for mean *state* valence, such that more negative valence predicted higher depressive symptoms at the next fortnightly report ($\beta = -0.364$, SE = 0.164, $t(134.38) = -2.217$, $p = .028$) – but mean *trait* valence, in contrast, did not predict significantly ($\beta = 0.779$, SE = 2.283, $t(140.24) = 0.341$, $p = .733$). This difference between trait and state effects means that when an individual's daily screen content is more negative than their own usual level, they tend to report higher depressive symptoms at the next assessment, while people who generally consume more negative content than others are not more depressed overall than others.

Further, we found a significant interaction between the daily means of valence and arousal (i.e., state valence and arousal) of consumed content (β = 0.160, SE = 0.078, t(66.03) = 2.044, p = .045; **see Figure 3A**). Inspection of the interaction revealed that the relationship between content valence and depressive symptoms is moderated by day-to-day fluctuations in both valence and arousal (**see Figure 3A**). The association between more negative daily content and higher depression is strongest on days when the overall content consumed is lower in arousal. As the day's mean arousal increases, this association becomes weaker. In contrast, the interaction between trait valence and trait arousal means (i.e., overall means across the observation period) was not significant (β = −1.810, SE = 1.598, t(134.35) = −1.133, p = .259; **see Figure 3B**). This indicates that the stable, overall valence and arousal of the content consumed do not significantly predict depression when accounting for day-to-day fluctuations (state-level effects), but that the day-to-day valence fluctuations in screen content do predict the next self-reported depression score, and that low arousal, low valence screen content has the strongest negative effect.

The random-effects estimates for participants (**Table 1, bottom**) further indicated considerable between-participant variability in both baseline depressive symptoms levels (intercept SD = 11.38 [on the 0-60 CES-D scale]) and the sensitivity of depressive symptoms to daily affective fluctuations (e.g., state-valence slope SD = 1.74), suggesting that individuals differ markedly in how strongly their mood responds to changes in the valence and arousal of consumed content. In magnitude, the heterogeneity was substantial: the SD of baseline depression was nearly as large as the sample-wide mean (14.02), and variability in the state-valence slope was several times larger than the average fixed effect (–0.36), indicating wide differences in reactivity across participants.

*Table 1. Fixed and Random Effects of the Linear Mixed Model Predicting Depressive Symptoms (i.e., CES-D Total Score)*

| Effect | | | | | |
|---|---|---|---|---|---|
| **Fixed Effects** | **Estimate** | **Std. Error** | **df** | **t-value** | **p-value** |
| Intercept | **14.02*** | 1.10 | 143.02 | 12.70 | <0.001 |
| Valence Mean State | **-0.36*** | 0.16 | 134.38 | -2.22 | 0.028 |
| Arousal Mean State | 0.09 | 0.14 | 102.37 | 0.60 | 0.551 |
| Valence SD State | -0.01 | 0.15 | 133.63 | -0.04 | 0.971 |
| Arousal SD State | 0.16 | 0.17 | 128.03 | 0.95 | 0.345 |
| Valence Mean Trait | 0.78 | 2.28 | 140.24 | 0.34 | 0.733 |
| Arousal Mean Trait | -0.29 | 2.97 | 136.44 | -0.10 | 0.923 |
| Valence SD Trait | -0.66 | 1.72 | 139.55 | -0.38 | 0.702 |
| Arousal SD Trait | 1.24 | 2.30 | 139.95 | 0.54 | 0.591 |
| Valence Mean State * Arousal Mean State | **0.16*** | 0.08 | 66.03 | 2.06 | 0.044 |
| Valence SD State * Arousal SD State | 0.06 | 0.08 | 75.00 | 0.68 | 0.502 |
| Valence Mean Trait * Arousal Mean Trait | -1.78 | 1.64 | 133.69 | -1.09 | 0.279 |
| Valence SD Trait * Arousal SD Trait | -2.30 | 1.93 | 138.89 | -1.19 | 0.236 |

| **Random Effects (Participants)** | **Std. Dev** | **1** | **2** | **3** | **4** | **5** | **6** |
|---|---|---|---|---|---|---|---|
| Intercept | 11.38 | - | - | - | - | - | - |
| Valence Mean State | 1.74 | -0.22 | - | - | - | - | - |
| Arousal Mean State | 1.35 | 0.10 | -0.74 | - | - | - | - |
| Valence SD State | 1.63 | 0.00 | -0.02 | 0.15 | - | - | - |
| Arousal SD State | 1.75 | 0.02 | -0.09 | -0.44 | -0.52 | - | - |
| Valence Mean State * Arousal Mean State | 0.69 | 0.16 | -0.47 | 0.05 | -0.28 | 0.76 | - |
| Valence SD State * Arousal SD State | 0.77 | -0.25 | 0.09 | -0.01 | 0.38 | -0.11 | 0.20 |
| Residual | 5.71 | | | | | | |

Note: *$p < .05$; **$p < .01$.

The linear mixed-effects model predicting self-reported anxiety (**see Table 2 for full results**) revealed no significant main effects for either mean state valence ($\beta = 0.009$, SE = 0.18, t(132.90) = 0.054, p = .957) or mean trait valence ($\beta = -0.517$, SE = 2.44, t(138.78) = -0.212, p = .833). However, we again observed a significant interaction between the daily means of state valence and state arousal ($\beta = 0.220$, SE = 0.07, t(68.38) = 3.031, p = .003; **see Figure 3C**).

Inspection of this interaction showed that the relationship between valence and anxiety was moderated by daily arousal levels: on days when individuals consumed content with lower mean arousal, the association between negative valence and higher anxiety was stronger. In contrast, the interaction between trait valence and trait arousal means was not significant ($\beta = -1.568$, SE = 1.76, t(137.81) = -0.889, p = .376; **see Figure 3D**).

Together, these findings underscore the importance of state-level dynamics in predicting mental health outcomes. The day-to-day experience of consuming more negative valence content than usual is predictive of depressive symptoms, and less arousing and more negatively valenced content was a predictor of both depressive symptoms and anxiety. In contrast, stable, trait-like patterns of affective content consumption were not systematically related to differences in depressive symptoms and anxiety.

Similar to the depressive symptoms model, the random-effects estimates for the anxiety model (**Table 2, bottom**) (intercept SD = 11.97 [on the 20-80 STAI-T scale]; state-valence slope SD = 1.94; state-arousal slope SD = 1.92) reveal substantial heterogeneity across participants in both baseline anxiety and their responsiveness to daily shifts in affective content, underscoring individual differences in how digital experiences are associated with anxious affect

*Table 2. Fixed Effects of the Linear Mixed Model Predicting Anxiety (i.e., STAI-Trait Total Score)*

| Effect Fixed Effects | Estimate | Std. Error | df | t-value | p-value |
|---|---|---|---|---|---|
| Intercept | **42.57***  | 1.17 | 140.95 | 36.46 | <0.001 |
| Valence Mean State | 0.01 | 0.18 | 132.90 | 0.05 | 0.957 |
| Arousal Mean State | -0.33 | 0.18 | 103.54 | -1.80 | 0.075 |
| Valence SD State | -0.07 | 0.15 | 133.89 | -0.43 | 0.670 |
| Arousal SD State | 0.26 | 0.18 | 130.65 | 1.50 | 0.135 |
| Valence Mean Trait | -0.52 | 2.44 | 138.78 | -0.21 | 0.833 |
| Arousal Mean Trait | -1.95 | 3.19 | 137.25 | -0.61 | 0.542 |
| Valence SD Trait | -0.30 | 1.84 | 138.43 | -0.16 | 0.870 |
| Arousal SD Trait | 3.59 | 2.46 | 138.93 | 1.46 | 0.146 |
| Valence Mean State * Arousal Mean State | **0.22**** | 0.07 | 68.38 | 3.03 | 0.003 |
| Valence SD State * Arousal SD State | -0.02 | 0.07 | 89.91 | -0.26 | 0.793 |
| Valence Mean Trait * Arousal Mean Trait | -1.57 | 1.76 | 137.81 | -0.89 | 0.376 |
| Valence SD Trait * Arousal SD Trait | -2.10 | 2.07 | 137.74 | -1.02 | 0.311 |

| Random Effects (Participants) | Std. Dev | 1 | 2 | 3 | 4 | 5 | 6 |
|---|---|---|---|---|---|---|---|
| Intercept | 11.97 | - | - | - | - | - | - |
| Valence Mean State | 1.94 | -0.15 | - | - | - | - | - |
| Arousal Mean State | 1.92 | 0.09 | -0.70 | - | - | - | - |
| Valence SD State | 1.68 | 0.07 | -0.36 | 0.36 | - | - | - |
| Arousal SD State | 1.90 | 0.03 | 0.10 | -0.52 | -0.57 | - | - |
| Valence Mean State * Arousal Mean State | 0.63 | 0.15 | -0.53 | -0.02 | 0.10 | 0.43 | - |
| Valence SD State * Arousal SD State | 0.66 | -0.12 | -0.07 | -0.12 | 0.25 | 0.25 | 0.34 |
| Residual | 4.85 | | | | | | |

Note: **$p < .01$; ***$p < .001$

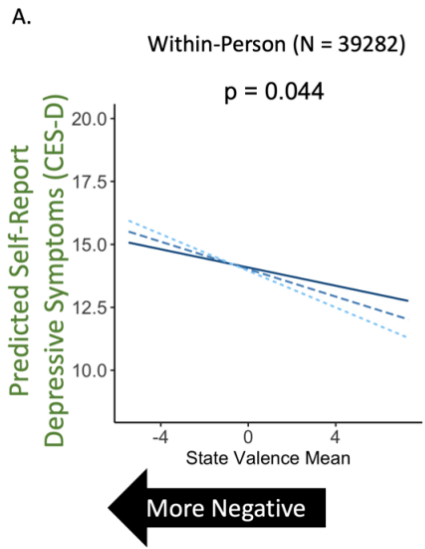

A. Within-Person (N = 39282), p = 0.044

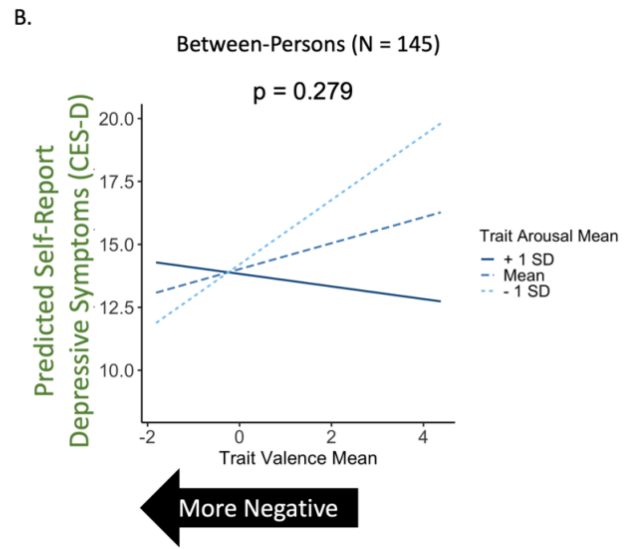

B. Between-Persons (N = 145), p = 0.279

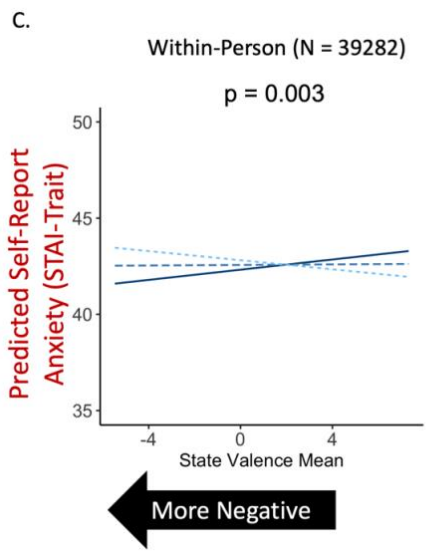

C. Within-Person (N = 39282), p = 0.003

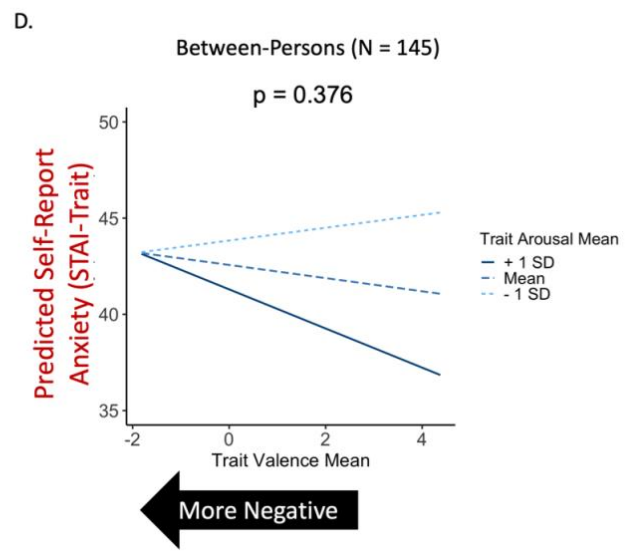

D. Between-Persons (N = 145), p = 0.376

**Figure 3. Interaction between State Valence Mean and State Arousal Mean on Self Report Depression and Anxiety. (A)** Interaction between *state* valence mean and *state* arousal mean predicting self-reported depressive symptoms (CES-D); the same interaction is shown for trait means in **(B)**. **(C)** Interaction between state valence mean and state arousal mean predicting self-reported anxiety (STAI-Trait); the same interaction is shown for trait means in **(D)**. Predicted depressive symptoms and anxiety scores are plotted against mean valence at varying levels of arousal (ranging from -1 SD [lightest blue] to +1 SD [darkest blue] from the mean), based on linear mixed-effects models. Models included random intercepts and random slopes for state-level valence and arousal across participants, capturing both individual differences in baseline affect and the effects of valence and arousal. Fixed effects for trait valence, trait arousal, and their standard deviations were also included. Significant interactions in **(A)** and **(C)** indicate that the combination of negative valence and low arousal is associated with higher depressive symptoms and anxiety scores, respectively. Nonsignificant interactions are displayed in **(B)** and **(D)**. *$p < .05$, **$p < 0.01$, N.S. = not significant.

**Discussion**

This study introduces the concept of the *Affective Digital Footprint* as a comprehensive representation of the emotional content people are exposed to in their everyday digital lives, derived from continuous capture of smartphone screen content across all apps. Using over 111 million screenshots collected approximately every five seconds from 145 adults over the course of a year, we quantify the valence and arousal of on-screen content and examine how both stable individual differences and day-to-day fluctuations in these affective features relate to mental health. Our findings show that dynamic, within-person deviations toward more negatively valenced content predict subsequent increases in depressive symptoms, and that day-to-day interactions between negative valence and low arousal predict higher levels of both depressive symptoms and anxiety. Importantly, it is short-term fluctuations in the affective qualities of consumed content, rather than differences in overall or average exposure, that are systematically associated with mental health outcomes across the year.

We found significant individual variability, with 58% of the variance in valence and arousal reflecting within-participant, day-to-day variability, while between-participant differences only accounted for approximately 42% of the variance. These results emphasize the importance of short-term affective dynamics in shaping mental health, pointing to the need for targeted interventions that address day-to-day variability in digital content consumption, and avoid self-reinforcing feedback loops of negatively valenced content.

***Interaction of Valence and Arousal in Predicting Depressive Symptoms and Anxiety***
Our models point to a significant interaction between state valence and arousal in predicting self-reported depressive symptoms and anxiety, indicating that the emotional impact of daily content depends on both its valence and arousal level. Consuming negatively valenced, low-arousal content predicted higher depressive symptoms and anxiety, whereas consuming positively valenced, high-arousal content predicted lower depressive symptoms but higher anxiety.

Our models revealed a significant valence and arousal interaction in predicting self-reported depressive symptoms and anxiety, such that negatively valenced, low-arousal content predicted higher depressive symptoms and anxiety, whereas positively valenced, high-arousal content predicted lower depressive symptoms but higher anxiety.

These findings replicate and extend the effects observed by Kelly and Sharot (2025), who demonstrated that valence plays a critical role in shaping mental health outcomes. Our work incorporates arousal as an additional dimension of content, demonstrating that the interaction between valence and arousal predicts depressive symptoms and anxiety in more nuanced ways. This supports theories of emotion that posit emotions are multidimensional constructs where valence and arousal jointly shape affective experiences (Russell, 1980; Feldman-Barrett, 2017; Feldman-Barrett et al., 2006). One possible explanation is that at lower levels of arousal, the valence of content becomes a clearer or more salient signal, making negative valence more impactful on the individual's emotional state. Engaging with content that is both negative in valence and low in arousal may therefore exacerbate feelings of depression or anxiety more than content that is negative but highly arousing. While there is bi-directional evidence evidence for the relationship between mental health and low affect of media consumption (Kelly & Sharot, 2025), an important open question for future work is the characteristic time scale of these loops - whether they unfold primarily over minutes and hours within smartphone use sessions or accumulate over days and weeks to shape more enduring changes in mental health.

**Implications**
These findings highlight the importance of state-level (vs. between-participant mean) fluctuations in valence and arousal as predictors of mental health outcomes, suggesting a more targeted approach than blanket screen time reduction strategies (Orben et al., 2020; Przybylski & Weinstein, 2017). Researchers, policymakers and developers should consider implementing and evaluating features that provide real-time feedback on users' affective content consumption. For example, operating systems could issue alerts when users show within-day accumulation of negatively valenced content, especially when it is also low-arousal, and offer tools to visualize daily affective exposures. Such interventions may enable users to proactively manage their content consumption in line with mental health goals (Kelly & Sharot, 2025).

At a broader level, public and policy discussions have largely framed screen use, especially among adolescents, as primarily a question of how much time is spent online, often positioning total screen time as a central driver of the youth mental health crisis (Twenge, 2018; Twenge et al., 2020; Stiglic & Viner, 2019; U.S. Surgeon General, 2023). Yet empirical findings on the relationship between screen time and mental health remain mixed, with studies reporting negative, null, and occasionally positive associations (Babic et al., 2017; Page et al., 2010; Granic et al., 2014; Odgers et al., 2018; Orben & Przybylski, 2019; Brusilovskiy et al., 2016; Sanders et al., 2024). This inconsistency suggests that duration of use is a coarse exposure metric that aggregates highly heterogeneous digital experiences and obscures psychologically meaningful differences in what individuals actually encounter during their time online. Across a year of continuous measurement, we find that short-term shifts toward more negatively valenced, low-arousal content, rather than overall amount of use, are associated with increases in depressive symptoms and anxiety. Although our observational design cannot establish causality, this pattern is consistent with self-reinforcing cycles of low mood and negative media consumption documented in prior work (Kelly & Sharot, 2025). Together, these findings argue for policy, research, and design efforts that move beyond screen time alone to explicitly account for the affective qualities of everyday digital content.

**Limitations and Future Directions**

Our study has a number of limitations. First, our sample may reflect selection bias, consisting of individuals willing to share their digital behavior. This, combined with our sample size (N = 145), may limit generalizability (Hargittai & Litt, 2011). Future studies should recruit more diverse samples across digital engagement levels, socioeconomic backgrounds, and age groups (Ellis et al., 2019). Second, our reliance on linear models may overlook more complex patterns in the relationship between features of individuals' Affective Digital Footprints of media content and mental health. Future research could employ nonlinear mixed-effects models or machine learning approaches to uncover more complicated temporal patterns and individual differences that may influence or reflect individuals' mental health. Third, while the Deep Affect model provides valuable insights into the content appearing on individuals' smartphone screens, the model may not fully capture affective nuances in textual or context-dependent content. Future work should integrate multimodal analysis, combining natural language processing and video analysis to provide a more comprehensive assessment of digital media's emotional content, including the possibility of tracking discrete specific emotions (e.g., despair) rather than mapping emotions into the valence-arousal circumplex. Finally, our observational design limits causal inference. Experimental studies manipulating digital content exposure could help establish causality and identify potential intervention strategies.

**Conclusion**

In summary, we combine psychological theory with computational analysis to better understand the impact of digital media on mental health. Our findings reveal that the valence and arousal of digital content—their *Affective Digital Screenshot*—significantly predict depressive symptoms and anxiety, highlighting the importance of capturing and describing the actual content that people engage with during their screentime. The insights obtained here through examination of the affective content of individuals' screens over a full year suggest the possibility of more nuanced strategies that go beyond merely reducing screen time, instead focusing on the affective qualities of content to improve mental health outcomes through informed, user-centered engagement.

**Methods**

**Participants**

Data examined here were provided by 145 adults recruited via Qualtrics to participate in a study as part of the Human Screenome Project (Mean Age = 45.4 years, SD =12.2 years ; Female = 51.4%, White = 46.6%, Black = 17.6%, Asian = 6.7%, Hispanic = 4.3%, Native American = 2.4%, Multiracial = 22.4%). Following our interest in how screen content and mental health change together over time, we included participants in this analysis that completed a substantial portion of the longitudinal protocol, providing self-report data and screenshots on 15 or more two-week periods (mean = 24.06 fortnights, sd = 2.99 fortnights) over the course of one year.

**Procedure**

After enrolling in the study and completing informed consent and HIPAA authorization for data collection as part of an IRB approved protocol at Stanford University, participants downloaded the Screenomics software (Kim et al., 2025) onto their smartphones, which then unobtrusively captured screenshots every five seconds when the device was in use. These screenshots, along with associated metadata (e.g., timestamp), were encrypted and securely transferred to privacy-protected research servers for subsequent analysis. Participants engaged in regular digital media usage over a one-year period using their personal smartphones. The Screenomics paradigm was employed to collect comprehensive records of participants' digital interactions. These 145 participants contributed a total of 111,270,108 screenshots during their study participation, daily average of 772,709 screenshots per person (range: [4,204 to 2,471,180]). Parallel to the screenshot collection, participants completed self-reports about their depressive symptoms and anxiety severity at fortnightly intervals via the Qualtrics online survey platform (Mean assessment interval = 15.3 days). The full Screenomics paradigm is described in more detail in (Brinberg et al., 2023; Ram et al., 2020; Reeves et al., 2021).

**Measures**

**Self-Report Assessments of Mental Health**

Throughout the year-long study, participants completed fortnightly self-reports (mean = 24.06) about their mental health via web-questionnaire, including measures of depression and anxiety.

*Depressive symptoms were* assessed using the Center for Epidemiologic Studies Depression Scale (CES-D; Radloff, 1977), a 20-item scale widely used as a measure of depressive symptoms. At each assessment, individuals indicated the frequency with which they experienced depressive symptoms over the past week (e.g., "*I felt that I could not shake off the blues even with help from my family or friends*" and "*I thought my life had been a failure*") on a 4-point Likert scale. Total scores computed as the sum of responses to the 20 items range from 0 to 60, with higher scores reflecting greater symptom severity.

*Anxiety* was evaluated using the State-Trait Anxiety Inventory-Trait (STAI-T; Spielberger et al., 1983), the trait subscale of the STAI, which assesses a person's general tendency to experience anxiety over time (in the previous 2-weeks). The STAI-T consists of 20 items, each rated on a 4-point Likert scale (1 = Almost Never, 4 = Almost Always), including: "*I am tense*", "*I am worried*", "*I feel calm*" (reverse-scored) and "*I feel secure*" (reverse-scored). Scores range from 20 to 80, with higher scores indicating greater trait anxiety. A license to use the STAI-T scale was obtained from Mind Garden, Inc.

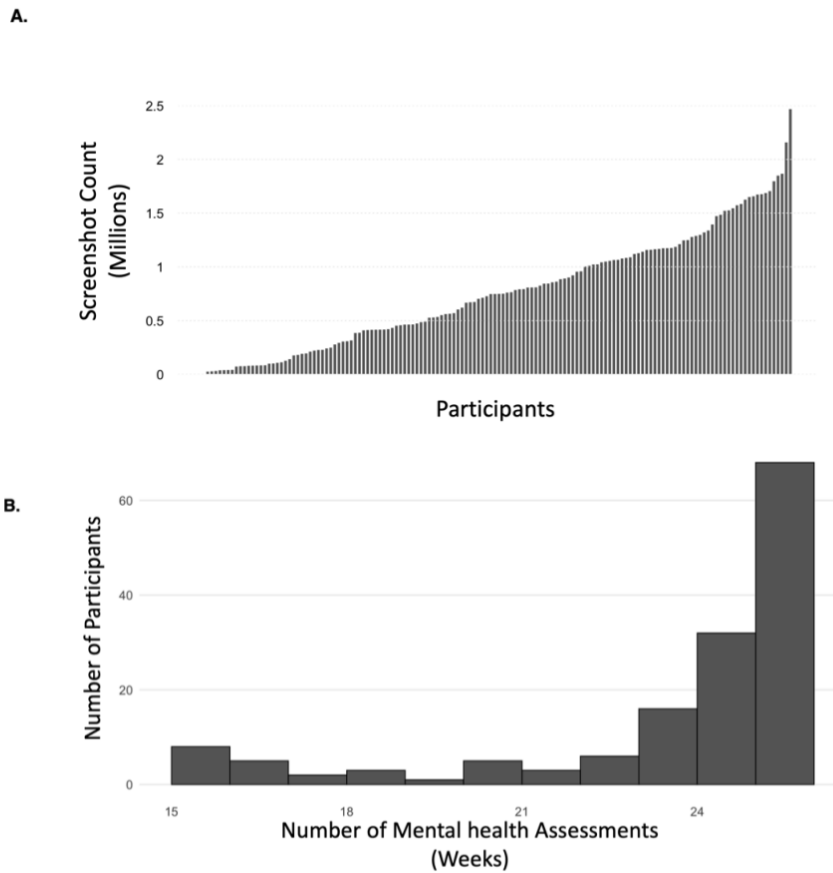

**Figure 4. Distributions of screenshot and mental health assessment frequency across participants. (A)** Distribution of screenshot counts per participant. Participants contributed a median of 751,692 screenshots (range: 4,204 to 2,471,180), reflecting substantial variability in smartphone engagement. **(B)** Distribution of mental health assessments completed per participant. Participants completed an average of 24.06 fortnightly assessments (SD = 2.99) over approximately one year.

**Daily Affective Content of Screenshots**
Ratings of the affective content of participants' screenshots was obtained using a computer vision model that scored each screenshot with respect to its affective valence and arousal. The Deep Affect model (Parry & Vuong, 2021) is a biologically inspired multistage deep neural network model that emulates the tiered structure of human visual processing to obtain valence and arousal scores for still images. The model uses three pre-trained convolutional neural networks (CNNs) based on VGG16 and ResNet architectures to extract high-level features related to facial expressions, scene contexts, and objects that appear in the images. These extracted features are subsequently processed by a deep feed-forward neural network with three hidden layers to generate numeric affect valence (1 = very negative to 9 very positive) and affect arousal (1 = calm to 9 = exciting) scores for each image. Originally trained on publicly available image repositories that include both images and average human ratings of affective valence and arousal, the model was later adapted by Rocklin et al. (2023) for use with smartphone screenshots. Their validation study showed that the model's ratings were consistent with human judgments of affective content in smartphone screenshots; congruence between human and model was ICC = .402 for valence (95% CI [.159, .577]) and ICC = .612 for arousal (95% CI [.448, .728]). Arousal showed good reliability with the model, while valence demonstrated fair but positive agreement. Conducted on a diverse set of screenshots spanning a wide range of digital media contexts, this validation confirmed the model's effectiveness in capturing the affective tone of participants' screen content.

### *Affective Digital Footprints*

We defined *Affective Digital Footprints* as the longitudinal, multidimensional patterns of valence and arousal in the digital media content individuals consume over a day. We quantified daily affective content consumption patterns using valence and arousal scores derived from all screenshots collected from a participant's smartphone on a given day. Specifically, each participant's *Affective Digital Footprints* was summarized using four metrics: (1) daily valence, calculated as the mean valence score of all screenshots captured on that day; (2) daily arousal, calculated as the mean arousal score of all screenshots captured on that day; (3) daily valence variability, calculated as the standard deviation of valence scores across screenshots within that day; and (4) daily arousal variability, calculated as the standard deviation of arousal scores across screenshots within that day. These daily metrics then allowed us to systematically examine both stable, trait-like patterns and dynamic, state-level fluctuations in the affective content participants consumed over the year.

## Data Analysis

Our analysis of individuals' *Digital Affective Footprints* and how they were related to fluctuations in mental health was situated in a multilevel modeling framework that accommodated the nested nature of the data (daily repeated measures nested within persons).

**RQ1: Examining Between-Participant Differences and Within-Participant Stability in Valence and Arousal Content Consumption (*Affective Digital Footprints*).** To statistically evaluate the metrics of participants' *Affective Digital Footprints* described above, we employed random intercept multilevel models, partitioning the total variance observed in the four features of the *Footprints* (mean valence, mean arousal, SD of valence, and SD of arousal) into between-person and within-person components. Between-person variance is the extent of stable, trait-like individual differences in participants' affective content consumption, and the within-person variance is the extent of state-like daily fluctuations in participants' affective content consumption.

We calculated the Intraclass Correlation Coefficient (ICC) for each of these four affective features using the formula:

$$ICC = \frac{\sigma_{u0}^2}{\sigma_{u0}^2 + \sigma_e^2}$$

where $\sigma_{u0}^2$ represents between-participant variance (variance of random intercepts) and $\sigma_e^2$ is within-participant variance (residual variance), as derived from a random intercept multilevel model. Higher ICC values indicate that the data can mostly be explained by stable individual differences, whereas lower ICC values indicate that the data can mostly be explained by within-participant variability across time.

Random intercept models were fitted separately for each affective feature using the lmer function from the lme4 package (Bates et al., 2015) in R. Confidence intervals (95%) for ICC estimates were obtained using bootstrapping with the bootMer function in lme4. Specifically, we generated 1,000 bootstrap samples by resampling the nested data with replacement, refitting the random intercept models for each bootstrap sample, calculating ICC values, and extracting the 2.5th and 97.5th percentiles of the resulting ICC distributions to obtain percentile-based confidence intervals.

To interpret the ICC estimates relative to RQ1, we assessed whether *Affective Digital Footprints* primarily reflected stable individual differences (higher ICC) or were predominantly shaped by daily fluctuations (lower ICC). Thus, the ICC values and associated confidence intervals provided direct evidence regarding the degree to which participants' *Affective Digital Footprints* represent stable, trait-like characteristics versus dynamic, state-like variability.

**RQ2: Association Between Depression and Anxiety from Affect of Content Consumed: Mixed-Effects Modeling.** After establishing that there were both substantial between-person and within-person differences in all four features of the *Affective Digital Footprints* (based on the ICCs above), we examined how these features differed and fluctuated in relation to fluctuations in individuals' mental health using multilevel models that accommodated the nested nature of the data (days nested within persons). Prior to analysis, each of the four time-varying features of the *Affective Digital Footprints* (i.e., mean valence and arousal and their SD) were z-scored across subjects and then separated into within-person and between-person components. Trait Valence Mean and Trait Arousal Mean scores were calculated for each participant as the within-person mean of all the daily valence and arousal scores, respectively. Trait Valence SD and Trait Arousal SD scores were calculated for each participant as the within-person mean of all their repeated measures of daily valence SD and arousal SD scores. State Valence Mean, State Arousal Mean, State Valence SD, and State Arousal SD scores were then calculated for each day as the deviation from the individual's Trait scores.

The within- and between-person associations of the *Affective Digital Footprint* features and mental health (depression and anxiety in separate models) were then examined using a two-level model of the form:

*Level 1: Within-Person Model*

At Level 1, we modelled daily depression and anxiety separately for individual *i* on day *t* as:

$$\begin{aligned}Depression_{ti} \text{ or } Anxiety_{ti} = {} & \beta_{0i} + \beta_{1i} State\ Valence\ Mean_{ti} + \beta_{2i} State\ Arousal\ Mean_{ti} \\ & + \beta_{3i} State\ Valence\ SD_{ti} + \beta_{4i} State\ Arousal\ SD_{ti} \\ & + \beta_{5i}(State\ Valence\ Mean_{ti} \times State\ Arousal\ Mean_{ti}) \\ & + \beta_{6i}(State\ Valence\ SD_{ti} \times State\ Arousal\ SD_{ti}) + e_{ti}\end{aligned}$$

The person-specific intercept $\beta_{0i}$ represents an individual's baseline level of mental health, the terms $\beta_1$-$\beta_6$ capture person-specific relations between the daily fluctuations in each of the four features of the *Affective Digital Footprints* and their interactions and the fluctuations in depression or anxiety, and the residual $e_{ti}$ are unexplained within-person fluctuations that are assumed normally distributed with mean zero and variance $\sigma_e^2$.

*Level 2: Between-Person Model*
The person-specific intercept and and slopes are then simultaneously modelled at level 2 as functions of the trait affective features:

$$\begin{aligned}\beta_{0i} = &\ \gamma_{00} + \gamma_{01} Trait\ Valence\ Mean_i + \gamma_{02} Trait\ Arousal\ Mean_i \\ &+ \gamma_{03} Trait\ Valence\ SD_i + \gamma_{04} Trait\ Arousal\ SD_i \\ &+ \gamma_{05}(Trait\ Valence\ Mean_i \times Trait\ Arousal\ Mean_i) \\ &+ \gamma_{06}(Trait\ Valence\ SD_i \times Trait\ Arousal\ SD_i) + u_{0i}\end{aligned}$$

$$\begin{aligned}\beta_{1i} &= \gamma_{10} + u_{1i} \\ \beta_{2i} &= \gamma_{20} + u_{2i} \\ \beta_{3i} &= \gamma_{30} + u_{3i} \\ \beta_{4i} &= \gamma_{40} + u_{4i} \\ \beta_{5i} &= \gamma_{50} + u_{5i} \\ \beta_{6i} &= \gamma_{60} + u_{6i}\end{aligned}$$

where the $\gamma$ parameters are sample-level relations between the trait-level features of the *Affective Digital Footprints* and mental health, and the $u$ terms are unexplained between-person differences that are assumed multivariate normal with variances $\sigma_{u0}^2, \sigma_{u1}^2, \sigma_{u2}^2, \sigma_{u3}^2, \sigma_{u4}^2, \sigma_{u5}^2, \sigma_{u6}^2$ and covariance $\sigma(u_i, u_j)$, for i ≠ j.

**Code availability**
Code is available via GitHub at https://github.com/CK-Cog/Affective_Digital_Footprints